\input{aipcheck}

\documentclass[
    ,final            
  ]
  {aipproc}

\layoutstyle{6x9}

\def\rrat {RRAT\,J1819--1458\,}

\newcommand{\XMM}{{\it XMM--Newton}\,}

\newcommand{\Swift}{{\it Swift}\,}

\def\ltsima{$\; \buildrel < \over \sim \;$}
\def\lsim{\lower.5ex\hbox{\ltsima}}
\def\loe{\lower.5ex\hbox{\ltsima}}
\def\gtsima{$\; \buildrel > \over \sim \;$}
\def\gsim{\lower.5ex\hbox{\gtsima}}
\def\goe{\lower.5ex\hbox{\gtsima}}

\def\ergs {erg\,s$^{-1}$}
\def\ergscm2 {erg\,s$^{-1}$cm$^{-2}$}
\def\ss {s\,s$^{-1}$}
\def\cm2 {cm$^{-2}$}

\begin{document}

\title{Rotating Radio Transients: X-ray observations}

\author{Nanda Rea}{address={University of Amsterdam, ``Anton Pannekoek'' Institute, Kruislaan 403, 1098~SJ, Amsterdam, NL}, altaddress={SRON Netherlands Institute for Space Research, Sorbonnelaan, 2, 3584~CA, Utrecht, NL} }

\begin{abstract}

Rotating RAdio Transients are a new class of neutron stars discovered
through the emission of radio bursts. Eleven sources are known up to
now, but population studies predict these objects to be more numerous
than the normal radio pulsar population. Multiwavelength observations
of these peculiar objects are in progress to disentangle their
spectral energy distribution, and then study in detail their
nature. In this review I report on the current state of the art
on these objects, and in particular on the results of new X--ray
observations.

\end{abstract}

\maketitle


\section{Introduction}

A new class of neutron stars displaying bursting behavior in the radio
band (RRATs, aka Rotating RAdio Transients) was recently discovered
(McLaughlin et al.~2006). These sources are all in the disk of the
Milky Way and are characterized by dispersed radio bursts with flux
densities (at a wavelength of 20~cm) ranging from 100 mJy to 4 Jy,
durations from 2 and 30 ms, and average intervals between repetition
from 4 minutes to 3 hours. Through a greatest common denominator study
of the arrival times of the bursts, periodicities for all of the RRATs
have been discovered. Furthermore, for the most prolific bursters
(RRAT\,1819--1458, RRAT\,1317--5759, and RRAT\,1913+1333), also period
derivatives ($\dot{P}$) and accurate positions ($<$ a few arcseconds)
have been measured. The timing solution derived from the radio bursts,
and their periods and period derivatives ranging from 0.4--7\,s and
10--14$\times 10^{-12}$\ss , respectively, indicate that they are
isolated neutron stars (NS). Nevertheless, the properties of their
emission are quite different from those of known isolated NSs, such as
e.g. radio pulsars (Lorimer \& Kramer~2005), magnetars (Woods
\& Thompson~2006), or X--ray dim isolated NSs (XDINS; van~Kerkwijk \&
Kaplan~2007).

Determining the nature of the emission from these objects, as well as
estimating how many RRATs are present in our Galaxy is of paramount
importance. The small duty cycle of the radio bursts (0.1--1\,s of
radio emission per day) makes the discovery of these sources rather
difficult, implying that there must be a large population of such
objects in the Galaxy. A detailed calculation shows that there may be
at least 5--10 times more objects of this class than canonical
rotational powered radio pulsars (McLaughlin et al.~2006; Lorimer et
al. in prep.).  Popov et al. (2006) show that the inferred birthrate
of RRATs is consistent with that of XDINS but not with magnetars.

There have been several suggestions put forward on the nature of this
new class of NSs. RRATs' radio bursting bahaviour might be similar to
the giant pulses sometimes observedin young radio pulsars, or to
``nulling'' phenomena observed as a turning on and off of the
pulsations. However, should be noted that the RRATs with a measured
$\dot{P}$, do not have high inferred values of magnetic field strength
at the light cylinder, implying that their emission mechanism is
different from that responsible for the ``giant pulses'' observed from
some pulsars (e.g. Knight et al.~2006). Furthermore, unlike most
nulling pulsars (e.g. Wang et al.~2007), we typically do not see more
than one pulse from the RRATs in succession (although with some
exceptions: McLaughlin et al. in prep.). Zhang et al.~(2006) suggest
that the RRATs may be neutron stars near the radio ``death line'',
however, the period derivatives measured for three RRATs do not place
them near canonical pulsar ``death lines'' (e.g. Chen \&
Ruderman~1993). Another intriguing possibility is that the sporadicity
of the RRATs is due to the presence of a circumstellar asteroid belt
(Cordes \& Shannon~2006; Li~2006) or a radiation belt such as seen in
planetary magnetospheres (Luo \& Melrose~2007). Or, perhaps, they are
transient X-ray magnetars, a particularly relevant suggestion given
the recent detection by Camilo et al.~(2006) of transient radio
pulsations from the anomalous X-ray pulsar XTE~J1810--197. A final
possibility is that they are similar objects to PSR~B0656+14, one of
three middle-aged pulsars (i.e. ``The Three Musketeers''; Becker \&
Truemper~1997) from which pulsed high-energy emission has been
detected (e.g. DeLuca et al.~2005).  Weltevrede et al.~(2006)
convincingly show that if PSR~B0656+14 were more distant its emission
properties would appear similar to those of the RRATs.


\begin{table}[t]
\begin{tabular}{cccccccccc}
\hline
\tablehead{1}{c}{b}{RRAT}
  & \tablehead{1}{c}{b}{Distance}
  & \tablehead{1}{c}{b}{P}
  & \tablehead{1}{c}{b}{Age}
  & \tablehead{1}{c}{b}{B-field}
  & \tablehead{1}{c}{b}{{\bf $\dot{E}$}}
  & \tablehead{1}{c}{b}{ S }
  & \tablehead{1}{c}{b}{X-ray Lum.}  \\

   & (kpc) & (s) & (Myr) & (Gauss) & (\ergs ) & (Jy) & (\ergs ) \\
\hline
1819--1458 & 3.6 & 4.26 & 0.117 & 5.0$\times10^{13}$ & 2.4$\times10^{32}$ & 3.6 & 3.9$\times 10^{33}$ \\
1317--5759 & 3.2 & 2.64 & 3.33 & 5.8$\times10^{12}$ & 0.3$\times10^{32}$ & 1.1 & $<7.5\times 10^{32}$ \\
1913+1333 & 5.7 & 0.92 & 1.86 & 2.7$\times10^{12}$ & 3.9$\times10^{32}$ & 0.6 & $<9.4\times10^{34}$ \\
\hline
\end{tabular}
\caption{Observed and derived parameters for RRAT\,J1819-1458, RRAT\,1317--5759, and RRAT\,1913+1333 (radio timing properties derived from McLaughlin et al.~2006).``S'' is the flux density at 1.4 GHz, and the X--ray luminosity is unabsorbed in the 0.3--5keV band (see text).}
\label{tab}
\end{table}


\section{X--ray observations}

\subsubsection{\rrat}

\rrat\, is one of the most prolific radio bursters, and shows the brightest 
radio bursts of any of the RRAT sources (see Tab.\,1 for details on
this source). Thanks to a serendipitous {\em Chandra} observation, we
detected X-ray emission from this source in a 30\,ks ACIS-I
observation toward the (unrelated) Galactic supernova remnant
G15.9+0.2 (Reynolds et al.~2006). The spectrum was described by an
absorbed blackbody, but the timing resolution was not sufficient to
allow a robust search for X-ray pulsations. We then re-observed the
source with \XMM\ on 2006 April 5th for 46~ks (see McLaughlin et
al.~2007 for details on the analysis). In this observation we
discovered X--ray pulsations (see Fig.\,1 left for a Z--squared plot)
and intriguing absorption features in the X-ray spectrum (McLaughlin
et al. 2007). This is the first detection of X-ray pulsations from any
of the RRAT sources. The X-ray pulsations were discovered at the same
periodicity derived from timing the arrival times of the radio bursts,
confirming the neutron star nature of these objects as well as the
technique used to time the radio bursts (McLaughlin et al. 2006).

In Fig.~1 (right), we present the X-ray profile over-imposed to the
radio observations of the bursts folded with the same ephemeris: the
radio bursts occurs exactly at the peak of the X-ray profile! The
X-ray pulse profile has a 0.3--5 keV pulsed fraction of 34$\pm$6\% and
can be well modeled as a single sinusoid, though there is a hint of
additional structure that more sensitive observations may reveal.  We
found no hint for a dependence of pulsed fraction on energy, although
our counts were not sufficient to significantly answer this point. We
found no evidence for any X-ray bursts or aperiodic variability on any
timescale.

\begin{figure}
\includegraphics[height=.4\textheight,width=6.3cm]{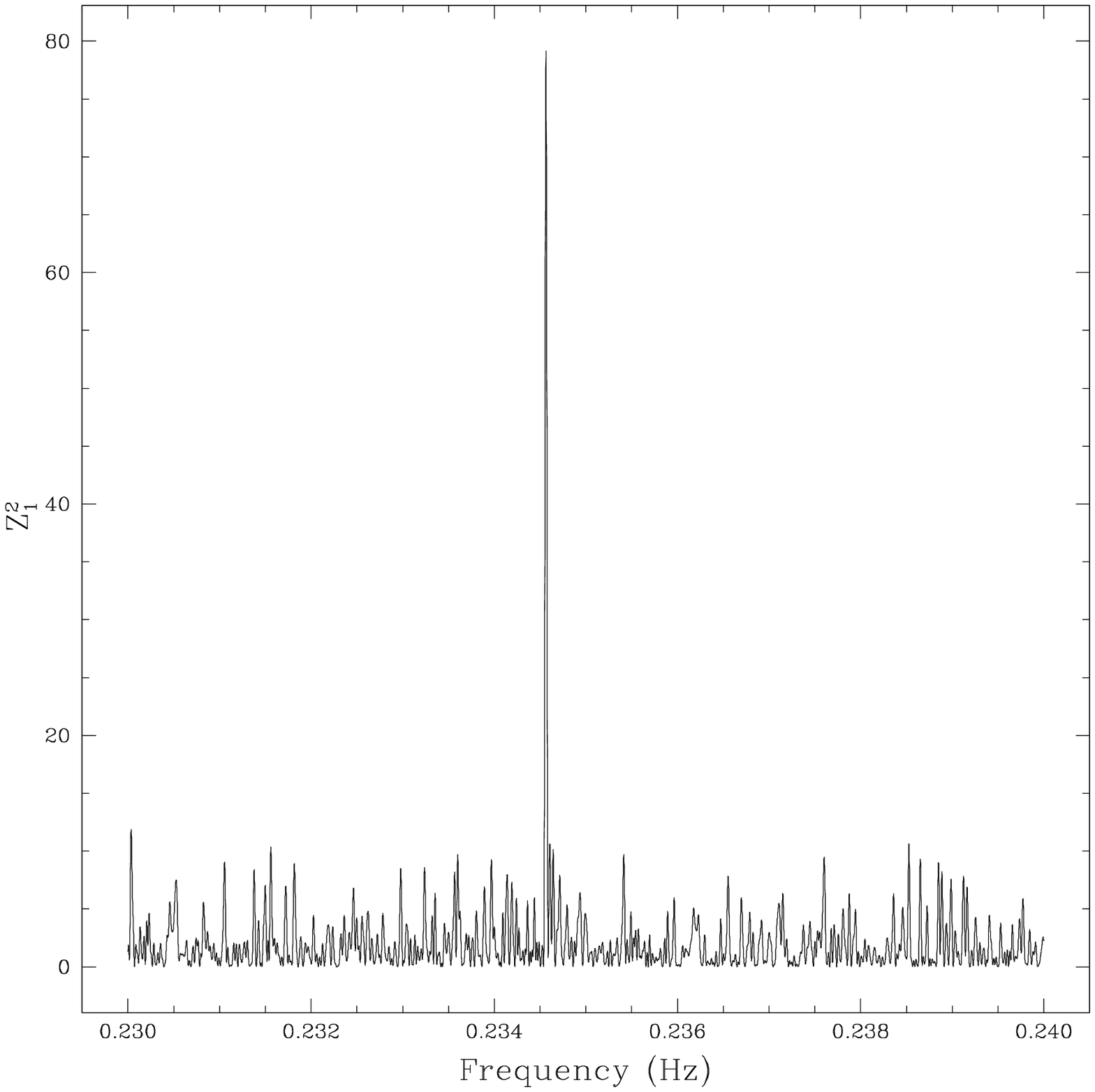}
\hspace{0.6cm}
\includegraphics[height=.4\textheight]{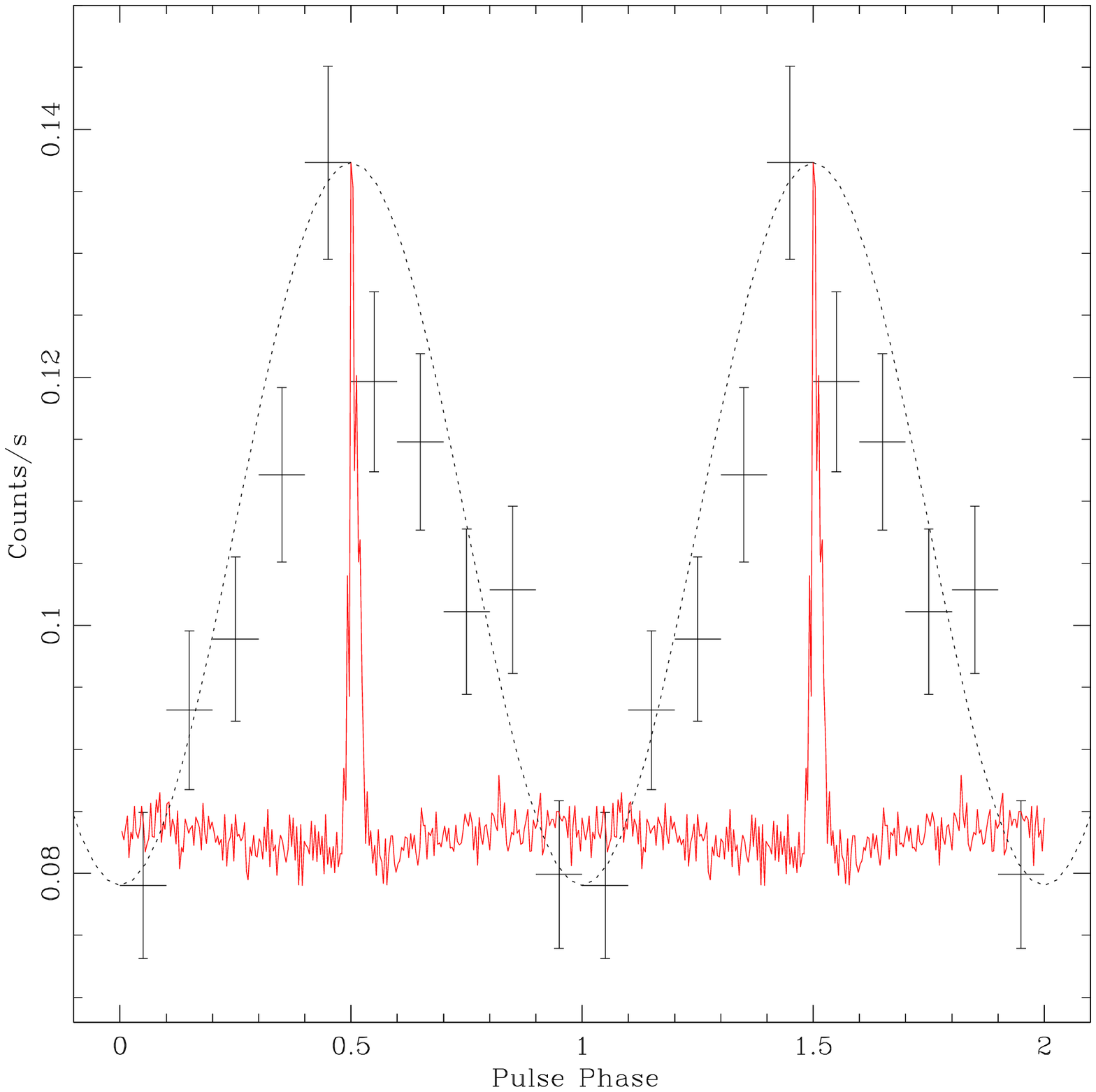}
\caption{{\em Left panel}: Z-squared probability of the X-ray period
  search. {\em Right panel}: X-ray pulse profile (0.3--5 keV) for
  \rrat. Dotted line shows the best-fit sinusoid. Radio profile formed
  from 114 pulses detected in 6 hr of observation at 1.4~GHz with the
  Parkes telescope in Australia is over-plotted. Both profiles have
  been folded using the radio ephemeris. In both plots, two cycles of
  pulse phase are shown.}
\end{figure}


The X-ray spectrum of \rrat\, showed even more surprises: a single
blackbody component did not fit the data because of the presence of
two strong features around 0.5 and 1\,keV (see Fig.\,2 left).  We
checked whether these might be due to calibration issues, to our
source and background extraction regions or to residual particle
flares and/or particles hitting the detector, and could reliably
exclude all of these.  We tentatively concluded that the 0.5-keV line
was not due to the source while to the Oxygen edge, hence we modeled
the spectrum excluding the 0.5--0.53~keV energy range, but more sensitive
observations are necessary to confirm this. Several models give
satisfactory results (giving our limited number of counts; see Tab.\,2
in McLaughlin et al.~2007 for details): we show in Fig.\,2 (right) the
blackbody (kT=0.14\,keV) plus a Gaussian (E$_G$=1.1\,keV) fit.  From
Monte Carlo simulations (see Rea et al.~2005, 2007a) we infer the
significance of the $\sim$1\,keV line to be 4$\sigma$. We tried to
perform a pulse phase spectroscopy dividing the observation in 2
phase intervals, but the limited number of counts did not allow us
neither to find significant spectral variability nor to give a strong
constrain on that. Furthermore, we found a hint for an additional
non--thermal component with $\Gamma\sim1$, dominating the spectrum
above 1.7\,keV, however, the addition of a further component was not
statistical significant given our number of counts, and the high
background dominating above 2\,keV did not allow us to disentangle
this issue.

The two main interpretations for the 1\,keV line is an atomic or
cyclotron line.  An atomic line could be due to the NS atmosphere or,
less probably, to a peculiar abundance in the ISM in the direction of
\rrat.  The structure which remains in in the residuals after fitting 
the broad line (see Fig.\,2 right), which in our data is not
statistically significant though, might be due to a blending of narrow
lines which are unresolved due to limited counts.  In the case of the
feature due to proton cyclotron resonant scattering, the magnetic
field inferred would be $2\times10^{14}$~G, broadly consistent with
that derived through radio timing of the bursts.  In addition, the
width and depth of the line are consistent with predictions for
proton-cyclotron absorption in highly magnetized neutron stars (Zane
et al.~2001). It is possible, although unlikely, that the feature is
the first cyclotron harmonic, with the 0.5\,keV fundamental coincident
with the depression in the spectrum that we have interpreted as due to
a mismodeling of the Oxygen edge. More counts are needed to
differentiate between these scenarios. Moreover, pulse phase
spectroscopy analysis is crucial for differentiating between the
atomic and cyclotron models, with phase variation expected in the
cyclotron hypothesis.  If the feature we detect is indeed due to
proton-cyclotron absorption, it provides an invaluable means of
testing the assumptions implicit in characteristic magnetic fields
derived through radio timing and an extremely valuable independent
measurement of the magnetic field of an isolated NS. We will soon
study these issues with a longer \XMM\, observation.

\subsubsection{RRAT\,J1317--5759}

We observed RRAT\,J1317--5759 with \XMM\, on 2006 July 16th, for a net
exposure time of 30\,ks. Data have been analysed following the same
procedures and softwares used in Rea et al.~(2007b). We find no X--ray
counterpart for this RRAT with an upper limit on the PN Small Window
counts of $<$0.012, which translates in an absorbed 0.3--5\,keV flux
of $<2.6\times10^{-14}$\ergscm2 , assuming an absorption (derived from
the source DM; Cordes \& Shannon~2002) of $N_H=5\times10^{21}$\cm2 and
a blackbody spectrum with kT=0.13\,keV (see Tab.\,1 for the
corresponding unabsorbed luminosity limit as well as for the distance
used).

\subsubsection{RRAT\,J1913+1333}

This RRAT has been observed with \Swift-XRT on 2005 November 20th for
an exposure time of 9.3\,ks. Data have been analysed as reported in
Rea et al.~(2007c). Also in this case we find no X--ray counterpart
for this RRAT with an upper limit on the \Swift--XRT Photon Counting
rate of $<$0.036 counts/s, which translates in an absorbed 0.3--5\,keV
flux of $<7.6\times10^{-13}$\ergscm2 , assuming an absorption of
$N_H=8\times10^{21}$\cm2 (again derived from the source DM; Cordes \&
Lazio~2002)and a blackbody spectrum with kT=0.13\,keV (see
Tab.\,1 for the corresponding unabsorbed luminosity limit as well as
for the distance used).

\begin{figure}
\includegraphics[height=.39\textheight,width=6.3cm]{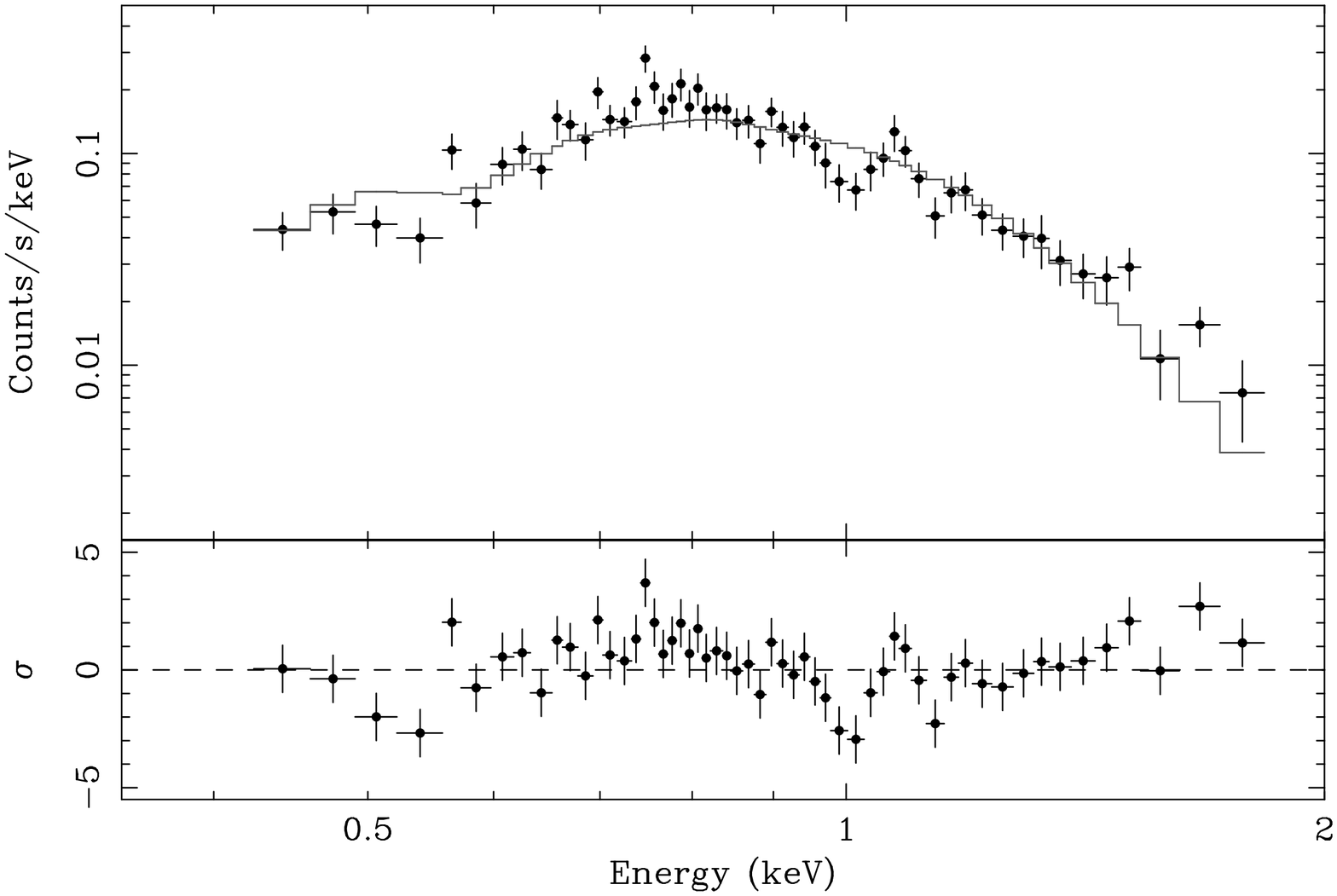}
\hspace{0.6cm}
\includegraphics[height=.4\textheight]{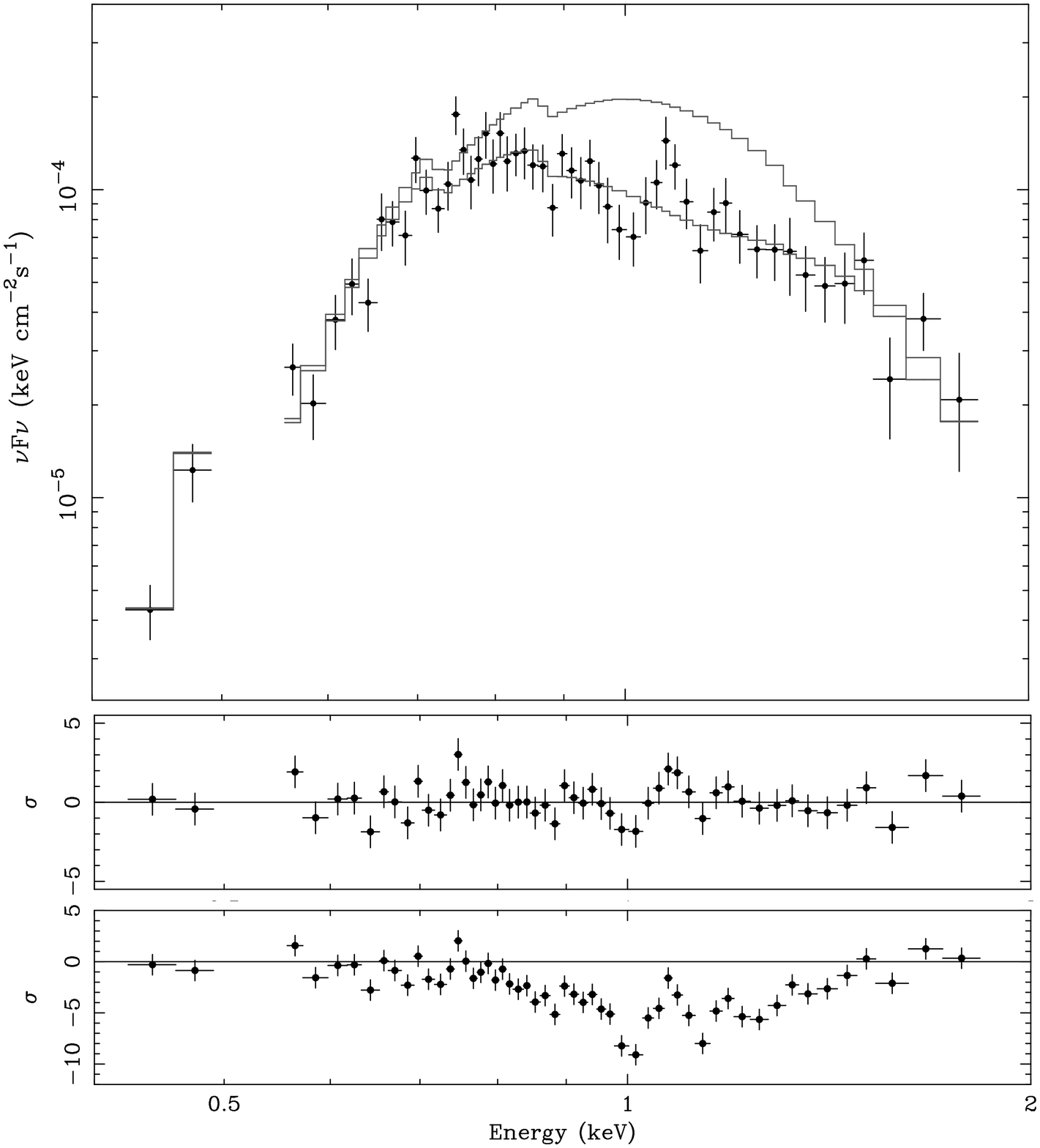}
\caption{{\em Left panel}: \XMM\, EPIC-PN spectrum of \rrat\, modeled with an
    absorbed blackbody. {\em Right panel}: Top: {\it XMM-Newton}
    EPIC-PN $\nu F \nu$ spectrum of \rrat , modeled with an absorbed
    blackbody and Gaussian line (see text). Top line shows the
    absorbed blackbody model alone. Middle: Residuals of the absorbed
    blackbody and Gaussian line model. Bottom: Residuals of the model
    without the inclusion of the Gaussian line.}
\end{figure}


\section{Discussion}

We presented in this review new X--ray observations of three RRATs,
\rrat , RRAT\,J1317--5759, and RRAT\,J1913+1333, showing that only
for the first one there is a robust detection (see Tab.\,1; Reynolds
et al.~2006; McLaughlin et al.~2007). However, present upper limits,
especially in the case of RRAT\,J1913+1333, are not very constraining,
and deeper X--ray observations are needed.

Our detection of X--ray periodicity at the radio period of \rrat\,
shows that the X-ray source we previously reported in Reynolds et
al. (2006) is undoubtedly RRAT's X--ray counterpart. The pulsed
fraction and sinusoidal pulse shape are similar to what is observed
for other middle-aged X-ray detected radio pulsars such as
PSR~B0656+14 (e.g. De Luca et al. 2005), which moreover has been
observed to have similar radio properties to the RRATs (Weltevrede et
al.~2006).

The thermal emission from \rrat\, is consistent with a cooling neutron
star. However, the temperature from our blackbody fit
(kT$\sim$0.14\,keV) appears slightly higher than temperatures derived
from blackbody fits for other neutron stars of similar ages (see
discussion by Reynolds et al.~2006).  Note that it is possible that
\rrat\, was born spinning at a sizable fraction of its present
period of 4.26~s.  In this case, as discussed by Reynolds et
al.~(2006), its characteristic age of 117\,kyr could be a considerable
overestimate, and the inferred temperature could be completely
consistent with its age.  Note that characteristic ages have been
shown to be misleading for several other pulsars (e.g. Gaensler \&
Frail 2000; Kramer et al.~2003).

Including \rrat, five over eight high-magnetic field radio pulsars
have now been detected at X-ray energies. In particular, two of them,
PSR~J1846--0258 and PSR~B1509--58, are bright non-thermal sources, as
expected given their young ages (less than 2\,yr). PSR~J1119--6127 is
a bright thermal X-ray emitter with unusual properties including a
large pulsed fraction and narrow pulse (Gonzalez et al. 2005).
PSR~J1718--3718, with magnetic field of $7\times10^{13}$~G, has been
detected at X-ray energies, but the faintness of the counterpart does
not allow detailed spectral modeling or a constraining limit on pulsed
fraction (Kaspi \& McLaughlin 2005). No X-ray emission has been
detected from the other high-B pulsars PSR\,J1814--1744, PSR\,B0154+61
or PSR\,J1847--0130, the latter has the highest inferred surface
dipole magnetic field ($9\times10^{13}$~G) measured to date for any
radio pulsar (McLaughlin et al.~2003). Radio pulsar X-ray emission
properties vary widely, even for objects with very similar spin-down
properties. Of course, also the radio emission properties of \rrat\,
are quite different from the radio emission properties of these other
high--B pulsars.

While the spectrum and luminosity of \rrat\, argue against a
relationship with magnetars, the recent detection of radio pulsations
from the transient magnetar XTE~J1810--197 (Camilo et al.~2006) raises
the interesting possibility that \rrat\, could be a transition object
between the pulsar and magnetar source classes.  The soft X-ray
spectrum does have a comparable temperature to the quiescent state of
XTE~J1810--197 ($kT \sim 0.15-0.18$ keV; Ibrahim et al.~2004; Gotthelf
et al.~2004).  However, the radio emission characteristics of these
two neutron stars are quite different. Would be very interesting to
search for RRAT--like bursts from this transient magnetar when it will
reach again the quiescent level.

While resonant cyclotron features are regularly observed from X-ray
binary systems (e.g. Truemper et al. 1978; Nakajima et al. 2006), the
detection of such features from isolated neutron stars is quite
unusual. Bignami et al. (2003) discovered two (maybe four; see also
Mori et al.~2005) harmonically spaced absorption lines from
1E~1207.4--5209, a radio-quiet X-ray pulsar with a 424-ms spin period
and timing-derived characteristic age and inferred surface dipole
magnetic field strength of $3\times10^{5}$~yr and $3\times10^{12}$~G,
respectively. 

Broad absorption lines, similar to those seen for \rrat, have been
observed for six out of seven XDINS (van Kerkwijk \& Kaplan 2007;
Haberl 2007). For most of these neutron stars, the lines can be
interpreted as due to neutral hydrogen transitions in highly
magnetized atmospheres.  Van Kerkwijk \& Kaplan~(2007) argue that the
transition energy is similar to the proton cyclotron energy for
magnetic fields of the order of a few $10^{13}$\,G.  The X--ray
spectrum of \rrat\, is very similar (although with a slightly hotter
blackbody temperature) to the XDINSs, although so far no convincing
evidence for radio bursts have been detected for any of those
thermally emitting neutron stars (Kondratiev et al.~2007~in prep.;
Burgay et al.~2007~in prep.; Rea et al.~2007d).

One outstanding question is why absorption lines of this kind, whether
due to the atmosphere or to cyclotron resonant scattering, have been
observed from only a handful of X-ray emitting isolated neutron
stars. The age of the neutron star could be one key factor. Young
objects are dominated by non-thermal emission, but older ones may be
too faint for X-rays to be detectable, making X-ray bright,
middle-aged pulsars the best candidates (as is the case of the XDINSs
and of \rrat ). Note, however, that no such absorption lines have been
found for the X-ray bright, middle-aged PSR\,B0656+14, despite deep
searches both with {\it Chandra} (Marshall \& Schultz 2002) and {\it
XMM-Newton} (De Luca et al. 2005). Another factor might well be the
viewing angle.


\begin{theacknowledgments}
The author thanks all the collaborators to this work:
M.~A.~McLaughlin, B.~M.~Gaensler, S.~Chatterjee, F.~Camilo, M.~Kramer,
D.~R.~Lorimer, A.~G.~Lyne, G.~L.~Israel, A.~Possenti, M.~Burgay,
S.~Reynolds and K.~Borkowski. NR is supported by an NWO {\tt Veni}
Fellowship and thanks the organizers of the meeting for the invitation
and the hospitality.

\end{theacknowledgments}







\end{document}